\input psfig.sty
\documentstyle[preprint,prl,aps,epsf]{revtex}         

\def\kpc{\,{\rm kpc}}

\def\km{\,{\rm km}}
\def\s{\,{\rm s}}

\def\kmps{\km\s^{-1}}

\def\dm{_{\rm DM}}
\def\rmsvel{\langle v^2\rangle^{1/2}}
\def\rmsveldm{\langle v^2\rangle_{\dm}^{1/2}}

\def\rmsveldmsun{\langle v^2\rangle_{\dm,\odot}^{1/2}}

\def\vc{V_c}
\def\vcinf{V_{c,\infty}}
\def\lsim{\raise0.3ex\hbox{$\;<$\kern-0.75em\raise-1.1ex\hbox{$\sim\;$}}}
\def\gsim{\raise0.3ex\hbox{$\;>$\kern-0.75em\raise-1.1ex\hbox{$\sim\;$}}}

\title{ Reply to Comment on "Dispersion Velocity of Galactic
 Dark Matter Particles" by Evans.}

\author{ 
R. Cowsik\footnote{cowsik, charu, pijush@iiap.ernet.in},
Charu Ratnam and P. Bhattacharjee}

\address{ Indian Institute of Astrophysics, Bangalore 560 034, INDIA.}

\begin{document}
\maketitle
\centerline{(Received:\hspace{5cm})}
\newpage
\tighten
In his comment on our {\it Letter}\cite{crb}, Evans\cite{evans} has 
expressed concern that the value of $\rmsveldm = 600\kmps$ derived
by us may not be valid in general because of our assumption of the 
Maxwellian form for the  distribution function (DF). 
He also concludes that there are undetected
numerical errors in our work. Neither of his points is valid. 
Below, we 
argue that our results are quite general and are only weakly 
sensitive to the precise form of the DF assumed in the analysis. In 
addition, we show that Evans' worries about numerical errors 
are unfounded. 

We have based our analysis on two of the simplest and most widely
used DFs, namely, (a) the Maxwellian and (b) the `lowered
isothermal' or `King model'\cite{bt}. The latter has the property 
that
both the spatial density and the velocity dispersion vanish at the
``tidal radius'' $r_t$, and $\rmsveldm$ depends on the 
galactocentric distance ($R$); it decreases from 
$\rmsveldm \sim \sqrt{3\sigma^2}$ at $R=0$ to $\rmsveldm=0$ at 
$R=r_t$, 
where $\sigma$ is the velocity parameter of the model\cite{bt}.    

\begin{figure}
\centerline{\psfig{figure=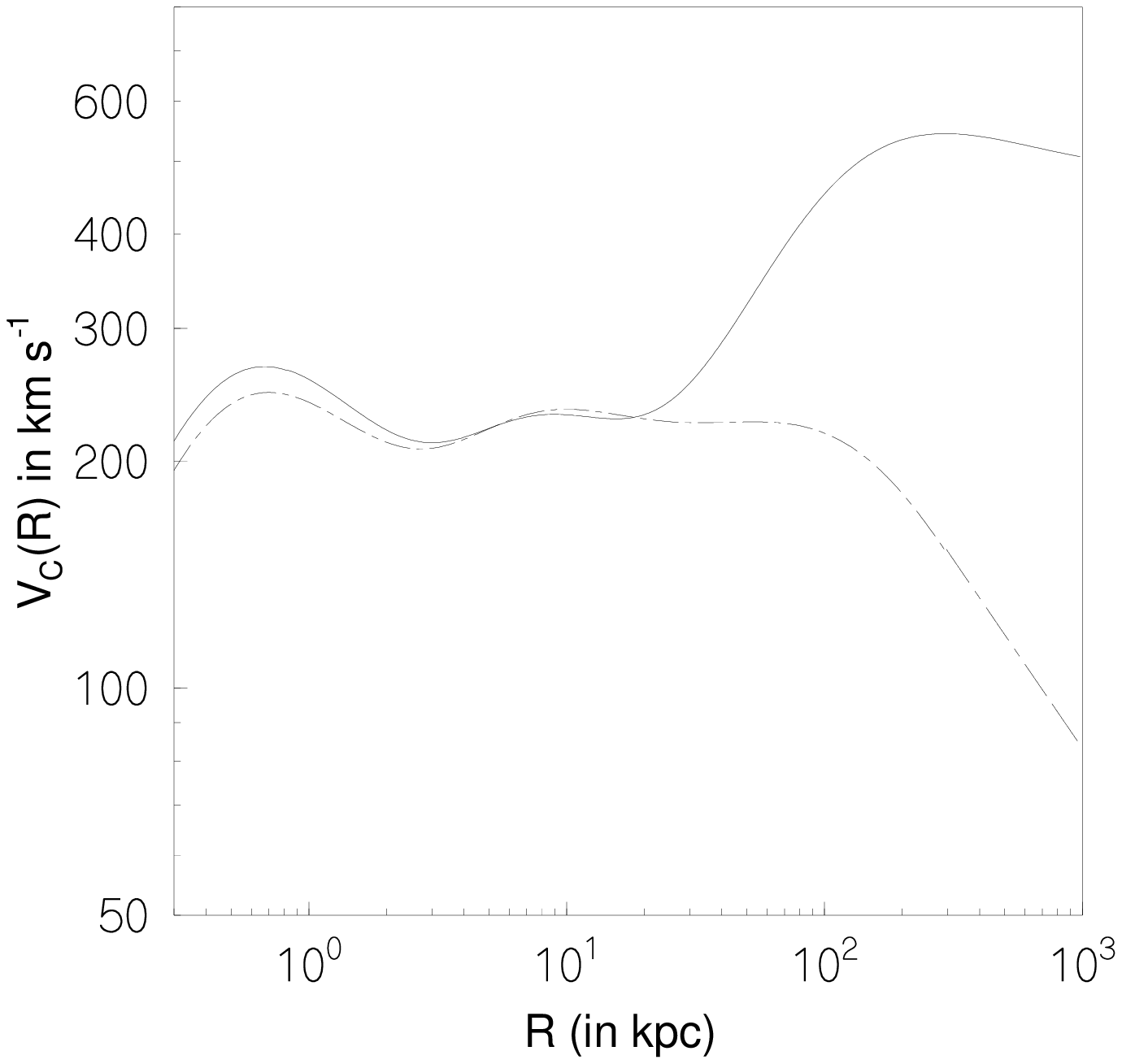,width=3.5in,height=2.5in}}
\caption{ Rotation curves of the Galaxy for the infinite isothermal
halo model (solid line) with $\rmsveldm = 600\kmps$ and for the 
lowered 
isothermal halo model (dashed line) for $r_t=300\kpc$ and 
$\sigma=330\kmps$ which corresponds to $\rmsveldmsun\sim570\kmps$.}
\end{figure}
Fig. 1 shows the rotation curves for the two DFs. 
With the King model the best fit to the rotation curve is obtained 
for $\sigma = 330\kmps$ which corresponds to the solar neighborhood 
value of dark matter velocity dispersion, 
$\rmsveldmsun \sim 570\kmps$,
not significantly different from $600\kmps$ for the Maxwellian DF 
(hence our comment in Ref. 18 of \cite{crb}). 
The robust nature of this result can be understood as follows: 
In the absence of streaming ($\langle v \rangle=0$), the leading 
moment, $\langle v^{2} \rangle$, appears as the pressure term in the 
Jeans equations (see Eq. 4-27 of \cite{bt}).
Thus for all
pressure supported halos the value of $\rmsveldmsun$ will not be 
significantly
different from the value of $\sim 600\kmps$ derived by us\cite{crb}.

Concerning the question of correctness of our numerical code,  
we note from Fig. 1 that for the  
case of the Maxwellian DF, even though the asymptotic relation for 
the circular velocity 
$\vcinf=\sqrt(\frac{2}{3})\,\rmsvel$ is 
{\it strictly valid only for single-component isothermal spheres}, 
 $\vc$ does indeed reach an asymptotic 
value implied by the above relation at $R\sim1000\kpc$. 
Moreover, even though it is difficult 
to anticipate fully the behavior of the solutions to the  
exponentially non-linear differential equations involved in 
the problem, 
the expectation of spherical symmetry at large distances is confirmed 
by our numerical calculations. 
At small $R$, the known visible matter dominates the gravitational 
potential, allowing us to have an approximate analytic expression 
for 
the dark matter density $\rho_{\dm}$(see Eq. (3) of Ref. \cite{crb}) 
which indicates that $\rho_{\dm}$ decreases with increasing 
$R$ more rapidly than for a single component isothermal sphere.
As already noted in \cite{crb}, we have cross-checked our numerical 
results
against these expectations, and in all cases we find excellent 
conformity to them.  

Finally we would like to emphasise the advantage
of using data on circular velocities, in deriving
the relevant halo phase space model parameters; 
$\vc^2$ directly yields $R\frac{\partial\phi}{\partial R}$, 
where $\phi$ is the total gravitational potential. 
In this context, note from Fig. 1 that while both Maxwellian and 
King forms of the DF predict 
essentially identical $\vc(R)$ up to $\sim 20\kpc$, the curves are 
very different at large $R$. This underscores the need to 
measure with greater precision $\vc$ as a function of R, 
especially at
large galactocentric distances, to fix the paramters describing the 
dark matter halo more exactly. 

In summary, we have shown that Evans'\cite{evans} criticisms  of our 
{\it Letter} \cite{crb}, are not valid and our result
of $\rmsveldm = 600\kmps$ is robust and stands unaffected.

\end{document}